\PassOptionsToPackage{table,dvipsnames,svgnames}{xcolor}

\documentclass[preprint]{vgtc}               

\usepackage{xcolor}

\makeatletter
\@ifpackageloaded{xcolor}{}{\usepackage{xcolor}}
\@ifundefined{color@NavyBlue}{\definecolor{NavyBlue}{rgb}{0,0,0.5}}{}
\makeatother

\graphicspath{{figures/}{pictures/}{images/}{./}} 

\usepackage{times}                     

\usepackage{tabu}                      
\usepackage{booktabs}                  
\usepackage{lipsum}                    
\usepackage{mwe}                       

\usepackage{mathptmx}                  
\usepackage{soul}
\usepackage{todonotes}

\usepackage{graphicx}
\usepackage{subfig}
\usepackage{subfig}

\usepackage{caption}

\PassOptionsToPackage{table,dvipsnames}{xcolor}
\usepackage{xcolor}

\usepackage{colortbl }

\usepackage{balance}

\usepackage[english]{babel}

\usepackage[markup=default]{changes}

\usepackage{array}
\newcolumntype{P}[1]{>{\centering\arraybackslash}p{#1}}

\newcommand\revision[1]{\textcolor{black}{#1}}

\onlineid{2004}

\vgtccategory{Research}

\vgtcinsertpkg

\preprinttext{Authors' version. To appear in an IEEE ISMAR 2025}

\title{ReachVox: Clutter-free Reachability Visualization for Robot Motion Planning in Virtual Reality}

\author{
Steffen Hauck\thanks{e-mail: steffen.hauck@stud.hs-coburg.de}\\
\parbox[c]{3.5cm}{\centering\scriptsize Coburg University of Applied Sciences and Arts}
\and Diar Abdlkarim\thanks{e-mail: d.abdlkarim@bham.ac.uk}\\
\parbox[c]{3.5cm}{\centering\scriptsize University of Birmingham}
\and John Dudley\thanks{e-mail: jjd50@cam.ac.uk}\\
\parbox[c]{3.5cm}{\centering\scriptsize University of Cambridge}
\and Per Ola Kristensson\thanks{e-mail: pok21@cam.ac.uk}\\
\parbox[c]{3.5cm}{\centering\scriptsize University of Cambridge}
\and Eyal Ofek\thanks{e-mail: e.ofek@bham.ac.uk}\\
\parbox[c]{3.5cm}{\centering\scriptsize University of Birmingham}
\and Jens Grubert\thanks{e-mail: jens.grubert@hs-coburg.de}\\
\parbox[c]{3.5cm}{\centering\scriptsize Coburg University of Applied Sciences and Arts}
}

 \teaser{
    \centering
    \includegraphics[width=\linewidth]{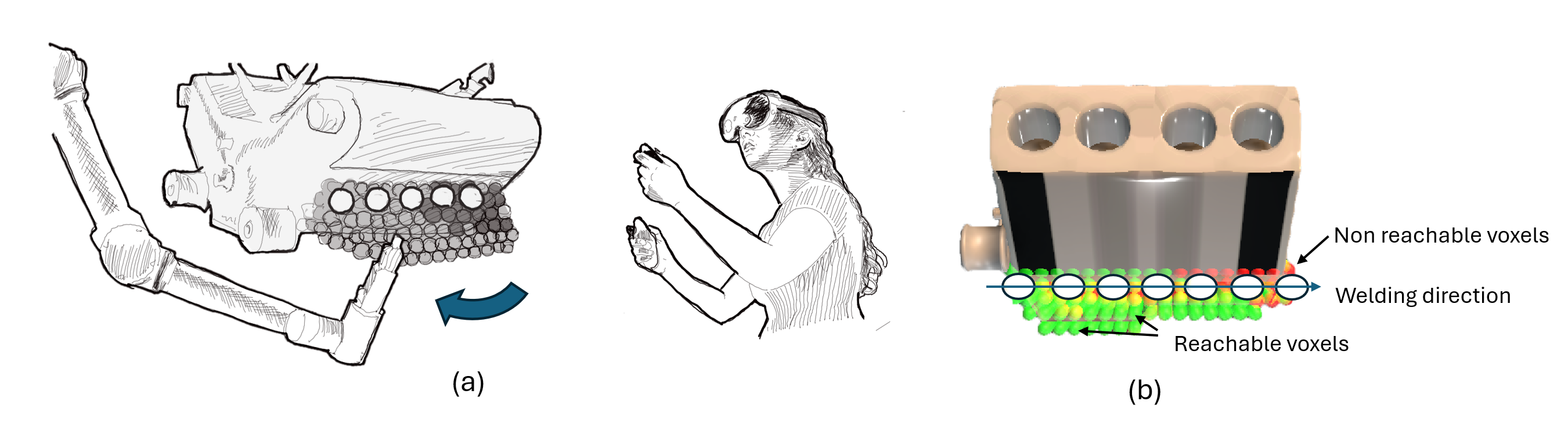}
    \caption{Remote Human-Robot-Collaboration: a) A remote operator needs to align the body of an engine so that a robot arm can access and weld it (a linear arrangement of white points represents the required welding locations). \textsc{ReachVox} is a visualization that can show global data, such as the areas that are reachable by the robot around the task area without visually cluttering the task space. Through this, the user controls the position and rotation of the engine, enabling her to align the engine efficiently. (b) A detailed view of the visualization around the welding area. The concentration of unreachable locations along the task area's right side indicates to the user the need to rotate the engine further toward the robot.}
    \label{fig:hero}
 }

\abstract{
   Human-Robot-Collaboration can enhance workflows by leveraging the mutual strengths of human operators and robots. Planning and understanding robot movements remain major challenges in this domain. This problem is prevalent in dynamic environments that might need constant robot motion path adaptation. In this paper, we investigate whether a minimalistic encoding of the reachability of a point near an object of interest, which we call \textsc{ReachVox}, can aid the collaboration between a remote operator and a robotic arm in VR. Through a user study (n=20), we indicate the strength of the visualization relative to a point-based reachability check-up.
} 


\begin{document}


\firstsection{Introduction}

\maketitle
Collaboration between human operators and robots can leverage the strengths of both.
Humans can better understand ad hoc situations and control them so that they are easily accessible by the robot. Remote operators who guide the robot from a safe location are an example of such collaboration. 
Their effectiveness is often controlled by their ability to understand the scene, its limitations, and the robot's capabilities and limitations.

In recent years, researchers have been looking at the use of immersive displays such as Virtual Reality (VR) and Augmented Reality (AR) to display human operators' information regarding robot status \cite{suzuki2022augmented, arevalo2021assisting}. For example, showing the intended motion of the robot \cite{gloumeau2025intuitive} can keep the operator safe by avoiding areas to which the robot is moving, or showing the accessible volume of a robot can help a designer to know where to place objects so that the robot can reach them.

\begin{figure}[!b]
  \centering
  \includegraphics[width=0.5\textwidth]{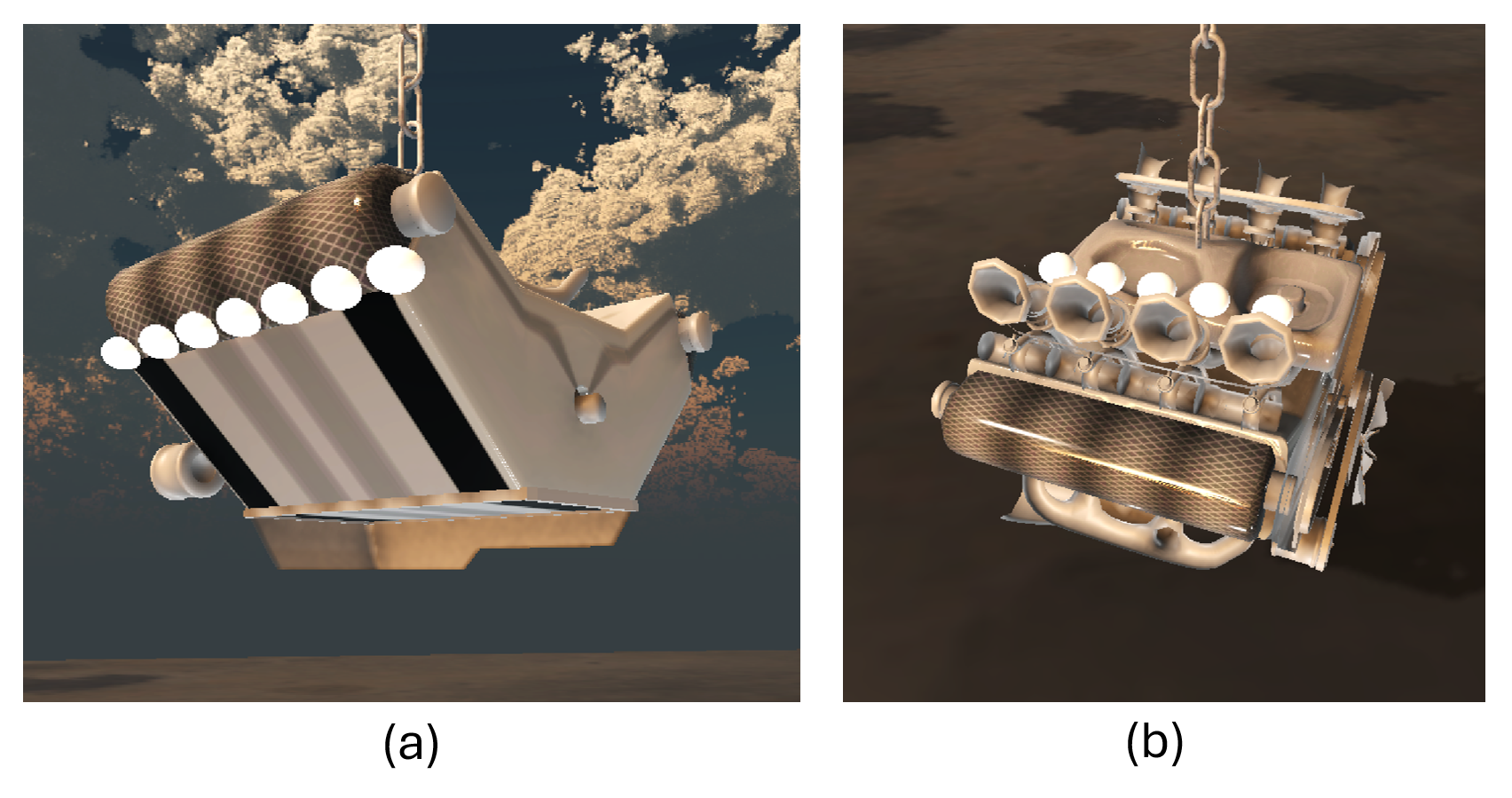}
  \caption{Example of an easy welding task (a) and a hard one, where the weld points lie inside an engine cavity, requiring greater maneuverability of the robot arm (b).}
  \label{fig:easyhard}
\end{figure}

We explore remote operation in ad hoc situations in which humans must collaborate with a robot arm to achieve a joint task. An example of such a use can be seen in Figure~\ref{fig:hero}. The robot needs to weld engine bodies that are accessed from a variety of directions.  The welding operation requires the robot to access a range of locations on the engine body in a given order. Some places, such as an area on the external part of the engine (as can be seen in Figure~\ref{fig:easyhard}, (a)) may be easy to access, while other areas that lie in a more concave part of the geometry (see Figure~\ref{fig:easyhard}, (b)) may require a lot of trial-and-error iterations until the operator can find how to orient the engine, so that the robot could access the whole task area and be able to weld it in one continuous motion.

This need requires visualizations that can concurrently present the results of algorithms, such as path planning and collision detection, over a large work area. In this way, an operator can see the effects of her decisions on the whole task space and better understand which operations are more likely to bring the robot closer to success. In this paper, we look at such a visualization within virtual reality. Instead of planning robot access to a single point at a time or along a single motion path, we found that it enables remote users to understand complex situations and aggregate visualization in a large area to make better decisions, and maximize the robot's accessibility to a workpiece.




This paper presents the following contributions.
\begin{enumerate}
\item A visualization of the reachability of robots in a task space aiming to reduce visual clutter is called \textsc{ReachVox}. The visualization enables \revision{novice} users to assess the probable success of a task, and an interactive update supports users to find suitable spatial configurations of a workpiece so that the robot arm has the best accessibility to it.
\item 
A user study (n=\revision{20}) that shows the strength of the proposed visualization relative to a point-based reachability check.
\end{enumerate}

\section{Related Work}

In situations where a robot path is not yet determined, e.g., in dynamic environments including obstacles, either the human or the robot needs to plan a path. Automatic robot motion path planning is a foundational problem in robotics, and even simplified versions of the motion planning problem are NP-hard \cite{lavalle2006planning, dai2022review}. Once the path is planned, various visualization techniques can be applied for the intended robot motion.

Pascher et al. \cite{pascher2023communicate} presented a scoping review of how to communicate robot motion intent (partly through AR) and classified techniques into motion, attention, state, and instruction but also noted the lack of a clear definition of robot motion intent. Suzuki et al. \cite{suzuki2022augmented} described four more areas in which AR can support human-robot-interaction beyond intent communication: programming, real-time control, safety, and increasing a robot's expressiveness.
With respect to intent communication, Andersen et al.~\cite{andersen2016projecting} proposed an object-aware projection technique to communicating robot intents through symbols (such as a warning sign) or indicating an area of interest that a robot will work on. Hietanen et al. \cite{hietanen2020ar} compared a projection-based work zone visualization with one presented on the Hololens in a user study (n=20) and indicated that users found the projection-based setup more plausible.  Hetherington et al. \cite{hetherington2021hey} used floor projection to visualize motions of a moving robot platform and Tsamis et al. \cite{tsamis2021intuitive} visualized the motion path of a robot arm mounted on a motion platform in an AR display for workers in a room-sized workspace shared with robots. Gruenefeld et al. \cite{gruenefeld2020mind}, compared three visualization techniques (path, preview, volume) in a user study (n=18) in which users should decide when a robot arm would enter a joint workspace. They found that the volume visualization was perceived as safest and required the least head rotations. Andronas et al. \cite{andronas2021multi} proposed to also visualize task instructions, recognized objects and warnings in addition to robot status and trajectory information. Arevalo et al. \cite{arevalo2021assisting} investigated awareness cues in an AR assisted teleoperation task to facilitate grasping tasks. For creating such visualizations, Lunding et al.~\cite{lunding2024robovisar} presented an AR authoring tool for in-situ robot visualizations. Amongst others, they implemented visualizations for showing the robot's path, state, force as well as safety zones. 

These visualizations have the potential to increase the user's awareness about the robot. Still, many works have been focusing on uncluttered environments or on minimizing the likelihood of robots and operators colliding in a joint task space.

With respect to visualizing motion and reachability constraints, e.g., due to obstacles in a robot's motion path, Rosen et al. \cite{rosen2020communicating} compared a 3D motion preview on a Hololens with a 2D display and a baseline (no visualization) in a user study with 32 participants. The participants were asked to decide whether a robot arm would collide with blocks in an environment. They found that the 3D preview outperformed the other techniques in terms of task completion time and error rate. 

 Gloumeau and Pettifer \cite{gloumeau2024exploring} compared two perspectives (front, third person) and three robot visualizations (opaque, transparent, and invisible) for VR-based robot teleoperation and found that users preferred third-person views and a transparent robot visualization due to mitigated occlusion issues. In a follow-up paper they \cite{gloumeau2025intuitive} compared three visualization techniques (complete, error and minimal feedback) in a user study (n=18) in the context of teleoperation of industrial robots. Their results indicated that the complete feedback technique outperformed the other techniques in both quantitative and qualitative metrics and led to a lower perceived task load.  
 
 Close to our work are capability \cite{zacharias2007capturing}, reachability \cite{makhal2018reuleaux}, and extended dexterity maps \cite{yao2023enhanced} that capture the reachability of points in space around the robot. We extend this idea to focus on the reachability around a workpiece with the goal of minimizing visual clutter and supporting human-robot collaboration in VR \revision{focusing on novice users}. 




\section{ReachVox}

\begin{figure}[tb]
  \centering
  \includegraphics[width=0.5\textwidth]{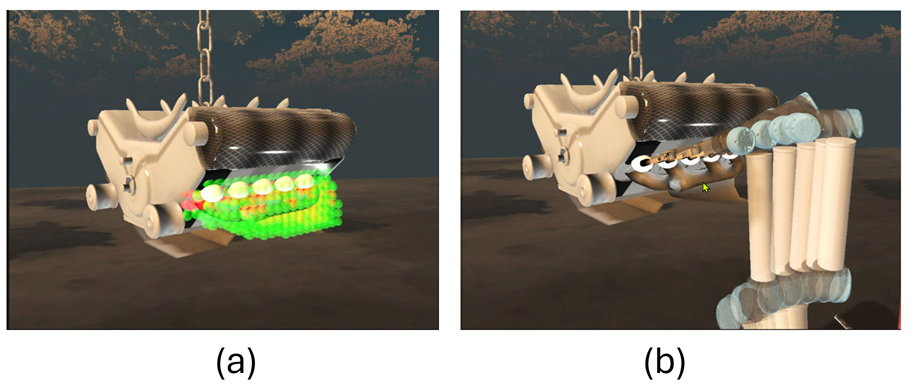}
  \caption{Reachable voxels around a workpiece (a) vs multiple previews (b) from the same point of view.}
  \label{fig:sbs}
\end{figure}

Although visualizing a robot's links and joints can be beneficial in understanding its spatial configuration when reaching a single point \cite{rosen2020communicating}, visualizing the reachability of multiple points on a surface or in a volume can quickly lead to visual clutter; see Figure \ref{fig:sbs}, (b). Complementary to previous work in AR and VR that focused on line or volume techniques once a single path is computed (e.g., \cite{gruenefeld2020mind}) and based on the idea of capability and reachability maps in robotics \cite{zacharias2007capturing, makhal2018reuleaux}, we introduce a visualization technique that aims at revealing the reachability of multiple points at once on a given 3D object without visually cluttering the task space. 


Our proposed visualization called ReachableVoxels (\textsc{ReachVox}) encodes the reachability status for the geometry around the task area (see Figures \ref{fig:hero}, (c) and \ref{fig:sbs}, (a)). The volume around the task is divided into cubic voxels. Each voxel on the surface of the geometry is represented as a colored semi-transparent sphere to display its reachability by the robot: A 'green' voxel indicates it is reachable by the robot end effector, without any collision detected between the robot's entire body and the scene geometry. In contrast, a red voxel signals that the simulation did not find a way for the robot to reach this voxel without colliding some part of the robot with the scene geometry.

This minimalistic visualization should enable the user to understand the global situation. For example, if only half of a welding path along an engine is accessible to the robot, it might be better to rotate the engine so that part is closer to the robot, with fewer possible obstructions. The visualization's feedback enables a user to try out different orientations and gradually progress toward one that maximizes the robot's accessibility to the task space. 

In general, computing the reachability for all voxels around the robot's maximum diameter is computationally very expensive (NP-hard) \cite{lavalle2006planning, dai2022review}. If the space required to be evaluated is constrained to limited volumes, such as the side of a workpiece that is going to be estimated, the computational costs can be reduced substantially. To this end, we intersect a voxel grid in the scene with the workpiece in question and only keep voxels around a specific threshold around the workpiece as active voxels. Still, for each active voxel, many possible robot configurations must be evaluated. As the focus of our work is not on real-time computations of robot motion, we utilized the following naive algorithm: We hierarchically iterate all robot joints with adjustable step sizes in the half-space between the robot and the workpiece. If the end-effector intersects with a voxel and no collision is detected with the scene geometry (including self-intersection of the robotic arm), the voxel's status is set to reachable (green) and red otherwise. In case a red voxel is visited again at a later stage, it can still turn green if a collision-free robot configuration occurs. Finally, the voxels are stored in \texttt{Dictionary<int[], bool>} where the integer array represents the voxel coordinates and the boolean represents if the voxel can be reached without collision (true) or not, and rendered as semitransparent spheres. The \textsc{ReachVox} code is available at \textit{$<$anonymized for review$>$}.

\section{User Study}

We conducted a user study to assess the benefit of \textsc{ReachVox} on a joint human-robot task: aligning a workpiece (in our case an engine) such that the robot could access target points on the workpiece (e.g., for spot welding).

\revision{We considered several different visualization options from related work \cite{rosen2020communicating, gruenefeld2020mind}. Using the terminology of Gruenefeld et al.~\cite{gruenefeld2020mind}, we considered \emph{path}, \emph{preview}, and \emph{volume} visualization. \emph{Path} visualizes the robot arm as interconnected line segments. However, while simple it is incomplete and does not address the aims of \textsc{ReachVox}, namely encoding whether or not a given point in space would lead to a collision between the robotic arm and the object of interest (including self-intersections). \emph{Volume} shows a translucent disc with colors indicating safe and unsafe regions given where the robot will position itself. Similar, to \emph{path} it has a different design objective compared to \textsc{ReachVox}.}

\revision{Finally, \emph{preview} consists of multiple   semitransparent poses of the robot along its path~\cite{rosen2020communicating}, see Figure \ref{fig:sbs}, (b). In a pilot study involving six participants (3 female, 3 male, mean age: 27.0 sd = 5.0), it became evident that the preview visualization technique was beneficial in illustrating the spatial extent occupied by the robot during motion tasks, thereby effectively supporting user decisions about safe positioning. However, feedback from the interviews revealed critical shortcomings for our specific collaborative engine-repair scenario. Participants consistently indicated that the volumetric cloud of semi-transparent robot poses characteristic of the preview visualization created visual clutter and occluded crucial areas of the task environment. This occlusion hindered clear identification of intersection events and negatively affected spatial understanding. Consequently, the results of this pilot study highlighted that the preview visualization, despite its value for safety-related spatial awareness, was unsuitable for inclusion in the main study due to its impaired visual clarity and increased cognitive demands for precise scene planning.}


Hence, we decided to use a baseline of a clear view of the scene and let the user guide the robot arm to the point they chose to see if any collisions were generated if the need would arise.

\begin{figure*}[tb]
  \centering
  \includegraphics[width=0.75\textwidth]{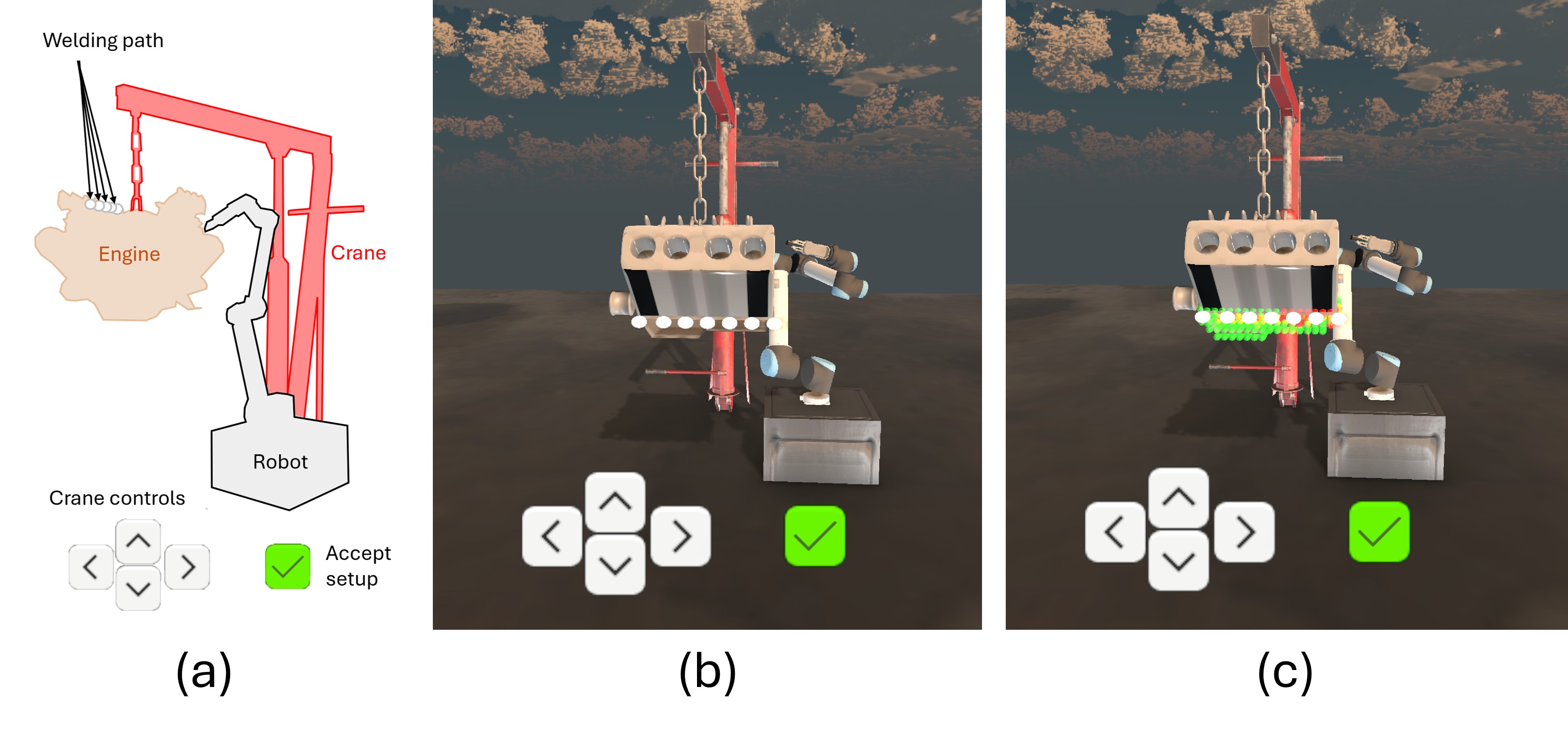}
  \caption{Overview over the workspace (a) and user study conditions: \textsc{base} (b) and \textsc{ReachVox} (c).}
  \label{fig:conditions} 
\end{figure*}

Figure \ref{fig:conditions}, (a) shows a typical task of the study. A participant can wear the HMD and hold the controllers in her hands. 
In front of her is a crane-supported car engine and a simulated UR10e industrial robot next to it.  Translation of the user in the scene is accomplished by the joystick on the left controller and rotation with the joystick of the right controller.   Motion sickness is mitigated through the dynamic field of view restriction \cite{fernandes2016combating}. 

A crane control panel is displayed in front of the participants (See Figure \ref{fig:conditions}). Participants can use the control panel to rotate the engine by pointing a controller to the left or right keys and pressing the controller button. At each key press, the crane rotates the engine 40\textdegree clockwise or counterclockwise. Another set of buttons allows the user to raise or lower the engine, in 4 steps, resulting in 36 possible spatial configurations for the engine. 

At any point during the study, a participant can grab the robot end effector (by pressing and holding the A button on the right controller) and drag it to a point of interest. The robot geometry follows the end effector using inverse kinematics, and Unity's physics engine is used to detect possible collisions between the robot arm and the engine geometry (including self-intersections of the robot arm). The task was to judge whether a robot could reach a set of contact points on the engine (indicated by white spheres).

The experiment was conducted as a within-subject design comparing two conditions: Our proposed visualization \textsc{ReachVox} and a baseline condition \textsc{base} with no additional visualizations, see Figure \ref{fig:conditions}. The robot's end effector could be moved in both conditions. Furthermore, in the conditions, we used two types of \textsc{difficulty}, see Figure \ref{fig:easyhard}. In \textsc{easy} the points to be reached were generally well visible and did not induce an intersection with the engine in multiple orientations. In \textsc{hard}, it was more challenging to judge in which of the 36 configurations intersection between the robot and engine would occur. These levels were empirically determined.

As dependent variables we collected task completion time, the time the robot was moved by participants, number of attempts to confirm the spatial configuration of the engine, and number of changes (rotations and height changes) made to the engine as objective measures. As subjective measures we used the after scenario questionnaire ASQ \cite{lewis1991psychometric} as a usability measure, the NASA TLX questionnaire \cite{hart1988development} (raw scores) as perceived workload measure, and an adopted robot trust questionnaire~\cite{charalambous2016development}. Finally, we collected preference ratings and comments from the participants. 

The order of the scenarios in the experiment was counterbalanced. In each condition, participants performed 8 trials (4 \textsc{easy} and 4 \textsc{hard} in randomized order).

\subsection{Apparatus}
\label{sec:Apparatus}

We used a virtual reality setup to simulate a remote operator that guides a robot on a new task. The system runs on a Windows 11 PC (NVIDIA Geforce RTX 4080, Intel Core i9-14900k, 64 GB RAM), running Unity game engine V. 2022.3.52f1, and using a Meta Quest Pro  HMD and its controllers. The simulated robot was a UR10e. For each of the 36 possible engine configurations, 2.6 million unique robot configurations were tested (around 25 seconds per configuration) to precompute the visualization. The step sizes for the joints (from base to end effector) were 30, 20, 12, 10, and 8 degrees. We skipped the sixth joint (rotation around the end-effector link), as it did not affect the final position of the tooltip of the robot arm. 


\subsection{Procedure}
Each participant signed a consent form and filled out a demographics questionnaire. 
The seated participants put on the HMD and took the controller in their hands. For each of the conditions (\textsc{ReachVox} and \textsc{Base}), participants first experienced a training scene in order to get familiar with the task and controls. During each of the conditions (which were conducted counter-balanced), they completed 8 trials using the crane controls (see \ref{sec:Apparatus}) until they were satisfied with the setup. They then pressed the accept button. If the setup resulted in a valid spatial configuration the next trial would start. If not they could try at most 8 times before the trial was classified as not successful and then the next trial would start. 

After each condition, the ASQ, TLX, robot trust questionnaires were filled out. After completing both conditions, participants were asked about their preferences and open comments. The participants were compensated with a voucher worth 15 € at the end and thanked for their participation. The study lasted ca. 60 minutes per participant.

\subsection{Participants}
We recruited a total of \revision{20} volunteers (\revision{13} male, \revision{7} female, mean age: \revision{29.2} sd: \revision{8.7}) all from a university population through convenience sampling. None of the participants reported any neurological or musculoskeletal conditions, and all had normal to corrected-to-normal vision. The participant average VR experience score was \revision{5.1} (sd \revision{2.0}) out of 7  (higher is better). We recorded a score of \revision{2.7} (sd = \revision{1.5}) and \revision{2.6} (sd = \revision{1.5}) out of 7 for experience with robots and experience with remote teleoperation systems, \revision{indicating that the participants were mostly  beginners in operating robotic systems}, as well as an average score of \revision{4.4} (sd = \revision{1.93}) out of 7 for experience with video games (again, higher indicates more experience).  

\subsection{Results}

We analyzed the influence of \textsc{visualization} and \textsc{difficulty} on the objective measures, using a two-way repeated measures analysis of variance (RM-ANOVA), with Greenhouse-Geiser correction if the sphericity assumption was violated. Post-hoc comparisons were conducted using pairwise t-tests with Tukey correction for multiple comparisons, controlling the error rate at an initial significance level of $\alpha = 0.05$. Effect sizes for the statistical tests were computed using Partial Eta Squared $\eta^2_p$. 
For the subjective measures, we used the Wilcoxon signed-rank test, with  Pearson's r as effect size measure. \revision{We did not observe any asymmetrical learning effects and there is no reason to suspect the experimental design induced any undue bias.} The data underlying the statistical analysis is available at \textit{$<$anonymized for review$>$}. 

\subsubsection{Task Completion Time}
Task completion time was measured from the start of a trial until the participant either successfully identified a valid spatial configuration or after 8 unsuccessful confirmations. The omnibus test indicated a main effect of \revision{\textsc{visualization}}, see Table \ref{tab:anova_tct}, were participants performed the task significantly faster in \textsc{ReachVox} (M = \revision{50.1}, Standard Error SE = \revision{5.0}), compared to \textsc{BASE} (M = \revision{62.0}, SE = \revision{6.1}). There was also an (expected) main effect of \revision{\textsc{difficulty}} with \textsc{Easy} being significantly faster (M = \revision{48.8}, SE = \revision{4.9}) than \textsc{Hard} (M = \revision{63.8}, SE = \revision{6.1}).

\begin{table}[!t]
    \centering
    \caption{RM-ANOVA results for task completion time. Significant findings are marked in gray. d$f_1$ = d$f_{effect}$ and d$f_2$ = d$f_{error}$. \revision{$\eta^2_p$: Partial Eta squared.}}
    \setlength{\tabcolsep}{6pt}

    \begin{tabular}{|c|c|c|c|c|c|}
        \hline
        \textbf{IV} & d$f_1$ & d$f_2$ & F & p & $\eta^2_p$ \\
        \hline
       Visualization ($V$) & 1 & \revision{19} & \revision{4.7} & \cellcolor{lightgray}$<0.05$ & \revision{0.2} \\
        \hline
        Difficulty ($D$) & 1 & \revision{19} & \revision{8.2} & \cellcolor{lightgray}$<0.05$ & \revision{0.3}\\
        \hline
        $V \times D$ & 1 & \revision{19} & \revision{2.0} & \revision{0.18} & \revision{0.09}\\
        \hline
    \end{tabular}

    \label{tab:anova_tct}
\end{table}

\begin{table}[!t]
    \centering
    \caption{RM-ANOVA results for robot movement time. \\
    }
    \setlength{\tabcolsep}{6pt}

    \begin{tabular}{|c|c|c|c|c|c|}
        \hline
        \textbf{IV} & d$f_1$ & d$f_2$ & F & p & $\eta^2_p$ \\
        \hline
       Visualization ($V$) & 1 & \revision{19} & \revision{4.1} & \revision{0.06} & \revision{0.2} \\
        \hline
        Difficulty ($D$) & 1 & \revision{19} & \revision{0.8} & \revision{0.37} & \revision{0.04}\\
        \hline
        $V \times D$ & 1 & \revision{19} & \revision{0} & \revision{0.93} & \revision{0}\\
        \hline
    \end{tabular}

    \label{tab:anova_robotmovetime}
\end{table}


\begin{figure}[!t]
  \centering
  \includegraphics[width=\linewidth]{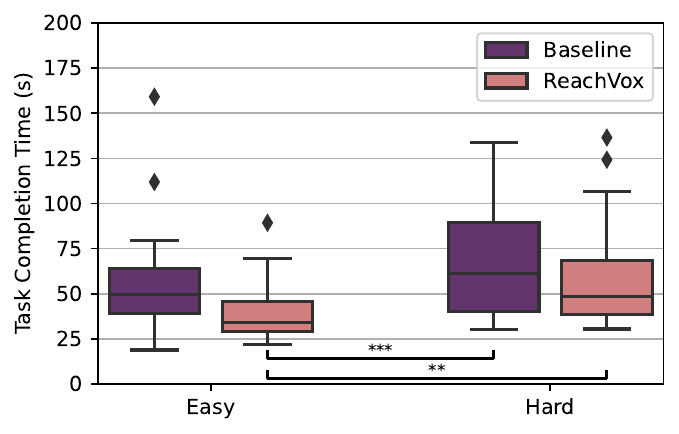}
  \caption{Task completion time in each of the conditions. $**: p < 0.01$. $***: p < 0.001$.}
  \label{fig:timetocomplete}
\end{figure}

\subsubsection{Robot Movement Time}
\begin{figure}[!t]
  \centering
  \includegraphics[width=\linewidth]{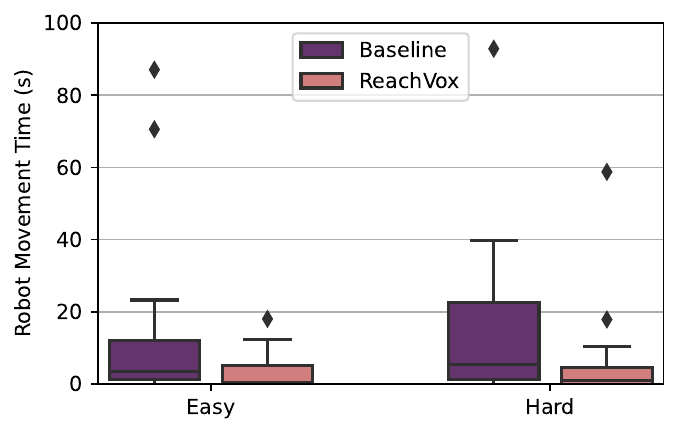}
  \caption{Robot movement time.}
  \label{fig:robotmovementtime}
\end{figure}

Robot movement time indicates the average time (in seconds) that the robot arm was moved by the participant during the trials. The omnibus test indicated no significant main or interaction effects, see Table \ref{tab:anova_robotmovetime} and Figure \ref{fig:robotmovementtime}. In \textsc{ReachVox}, 6 participants did not move the robot arm at all and in \textsc{base} one participant.





\subsubsection{Number of Attempts}

\begin{figure}[!b]
  \centering
  \includegraphics[width=\linewidth]{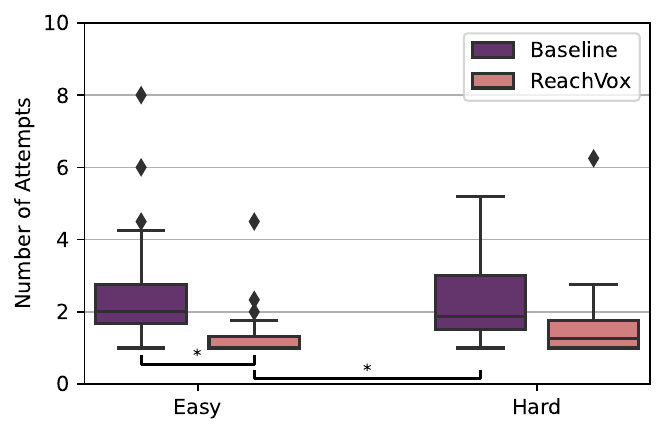}
  \caption{Number of attempts. $*: p < 0.05$.}
  \label{fig:attemptnumber}
\end{figure}

The number of attempts shows how often participants indicated that the spatial configuration would be correct. If the configuration was correct, the next trial would start, if it was not correct, the participants could try out a different configuration (up to 8 times in total per trial). The omnibus test indicated an overall main effect of \revision{\textsc{visualization}}, see Table \ref{tab:anova_numattempts} and Figure \ref{fig:attemptnumber}, where participants needed significantly fewer attempts in \textsc{ReachVox} (M = \revision{1.48}, SE = \revision{0.22}), compared to \textsc{BASE} (M = \revision{2.51}, SE = \revision{0.31}). 



\begin{table}[!b]
    \centering
    \caption{RM-ANOVA results for number of attempts. Significant findings are marked in gray.}
    \setlength{\tabcolsep}{6pt}

    \begin{tabular}{|c|c|c|c|c|c|}
        \hline
        \textbf{IV} & d$f_1$ & d$f_2$ & F & p & $\eta^2_p$ \\
        \hline
       Visualization ($V$) & 1 & \revision{19} & \revision{7.7} & \cellcolor{lightgray}\revision{0.01} & \revision{0.29} \\
        \hline
        Difficulty ($D$) & 1 & \revision{19} & \revision{0} & \revision{0.98} & \revision{0}\\
        \hline
        $V \times D$ & 1 & \revision{19} & \revision{1.75} & \revision{0.2} & \revision{0.08}\\
        \hline
    \end{tabular}

    \label{tab:anova_numattempts}
\end{table}


\subsubsection{Number of Changes}

\begin{figure}[!t]
  \centering
  \includegraphics[width=\linewidth]{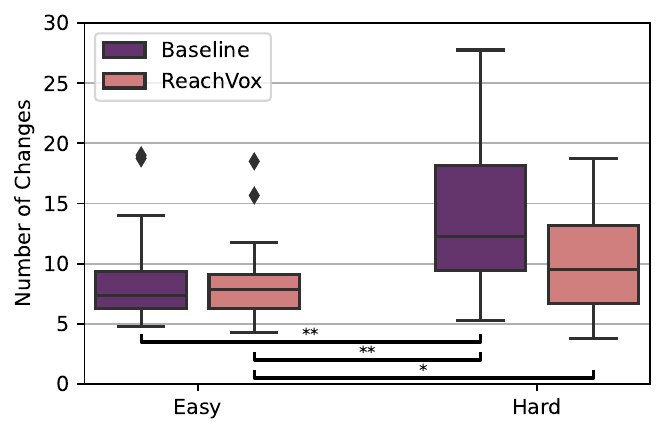}
  \caption{Average number of engine configuration changes performed by the participants. $*: p < 0.05$. $**: p < 0.01$.}
  \label{fig:numberofchanges}
\end{figure}

The number of changes indicates how often participants changed the rotation or height of the engine on average before the trial was completed. In total, there were 9 possible orientations $\times$ 4 possible heights = 36 possible spatial configurations for the engine. The omnibus test indicates \revision{a main effect of visualization}. Participants used significantly fewer changes in \textsc{ReachVox} (M = \revision{9.4}, SE = 0.8) compared to \textsc{base} (M = \revision{11.5}, SE = \revision{1.1}), see table \ref{tab:anova_numchanges} and figure \ref{fig:numberofchanges}. There was also a main effect of \revision{\textsc{difficulty}} with \textsc{easy} resulting on average in \revision{8.6} changes (SE = 0.7) compared to \textsc{hard} (M = 12.3, SE = \revision{1.1}) \revision{and an interaction effect (with post-hoc tests indicating a significantly lower number of changes for \textsc{ReachVox} compared to \textsc{base} in the \textsc{hard} condition ($p=0.02$, mean difference: $-3.6$, SE = $1.4$)}. 

\begin{table}[!t]
    \centering
    \caption{RM-ANOVA results for number of changes. Significant findings are marked in gray.}
    \setlength{\tabcolsep}{6pt}

    \begin{tabular}{|c|c|c|c|c|c|}
        \hline
        \textbf{IV} & d$f_1$ & d$f_2$ & F & p & $\eta^2_p$ \\
        \hline
       Visualization ($V$) & 1 & \revision{19}  & \revision{4.5} & \cellcolor{lightgray}\revision{0.046} & \revision{0.19} \\
        \hline
        Difficulty ($D$) & 1 & \revision{19} & \revision{20.3} & \cellcolor{lightgray}$<0.001$ & \revision{0.52} \\
        \hline
        $V \times D$ & 1 & \revision{19} & \revision{4.95} & \cellcolor{lightgray}\revision{0.038} & \revision{0.21}\\
        \hline
    \end{tabular}

    \label{tab:anova_numchanges}
\end{table}


\subsubsection{After Scenario Questionnaire}

\begin{figure}[!b]
  \centering
  \includegraphics[width=\linewidth]{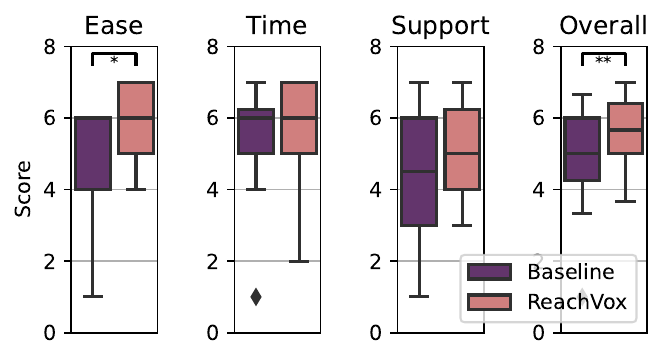}
  \caption{After Scenario Questionnaire ratings. $*: p < 0.05$. $**: p < 0.01$.}
  \label{fig:asqresults}
\end{figure}

The After Scenario Questionnaire (ASQ) \cite{lewis1991psychometric} encompasses three usability-related items that measure a user's satisfaction with a specific scenario or task. The overall score is the weighted average of the 3 subscores. Wilcoxon signed-rank tests indicated that \textsc{ReachVox} resulted in significant better ratings for \revision{the ease} subscore and the overall usability score, see Table \ref{tab:asq} and Figure \ref{fig:asqresults}.


\begin{table}[!htb]
    \centering
    \caption{Wilcoxon signed-rank test results for the ASQ for satisfaction with ease of task completion (Ease), with the time required (Time), with the support provided (Support) as well as the overall score (Overall). md: mean difference. Significant findings are marked in gray.}
    \setlength{\tabcolsep}{6pt}

    \begin{tabular}{|c|c|c|c|c|}
        \hline
        \textbf{Measure} & W & md & p & $r$ \\
           \hline
       Ease  & \revision{13.0} & -1.5 & \cellcolor{lightgray}$\revision{0.01}$ & \revision{-0.75} \\
        \hline
        Time  & \revision{21.0}  & -1.0 & \revision{0.16} & \revision{-0.46}\\
        \hline
        Support  & \revision{18.0} & \revision{-1.5} & \revision{0.056} & \revision{-0.60}\\
        \hline
        Overall  & \revision{24} & \revision{-1.0} & \cellcolor{lightgray}$<0.01$ & \revision{-0.75}\\
        \hline
    \end{tabular}

    \label{tab:asq}
\end{table}
    

\subsubsection{Robot Trust Questionnaire}
\begin{figure}[!t]
  \centering
  \includegraphics[width=\linewidth]{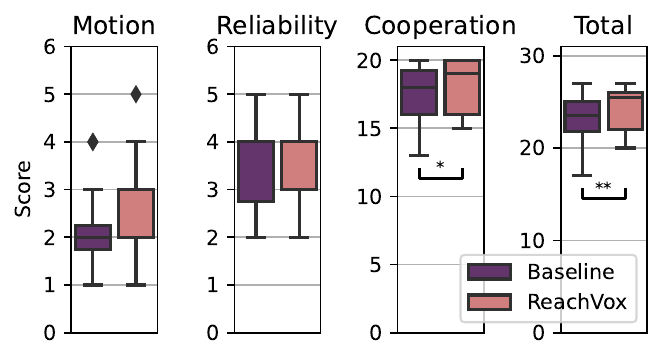}
  \caption{\revision{Trust scores for perceived robot's motion, perceived robot reliability, perceived safe cooperation, and the total trust score. $*: p < 0.05$. $**: p < 0.01$.}} 
  \label{fig:trustscore}
\end{figure}

For measuring the trust in human-robot-collaboration we adopted the robot trust questionnaire by Charalambous et al.~\cite{charalambous2016development} by removing three questions related to a gripper, which \revision{were} not relevant in our scenario. Wilcoxon signed-rank tests indicated that \textsc{ReachVox} resulted in significant higher perceived safe cooperation \revision{and total trust score} compared to \textsc{base}, see Table \ref{tab:trust} and Figure~\ref{fig:trustscore}.



\begin{table}[!t]
    \centering
    \caption{Wilcoxon signed-rank test results for the robot trust questionnaire for perceived robot's motion (Motion), perceived safe cooperation (Cooperation), perceived robot reliability (Reliability), and the total trust score (Total). md: mean difference. Significant findings are marked in gray.}
    \setlength{\tabcolsep}{6pt}

    \begin{tabular}{|c|c|c|c|c|}
        \hline
        \textbf{Measure} & W & md & p & $r$ \\
           \hline
       Motion  & \revision{10.5}  & \revision{-1.0} & \revision{0.32} & \revision{-0.42} \\        
        \hline
        Reliability  & \revision{20.0} & \revision{-0.5} & \revision{0.24} & \revision{-0.39}\\
        \hline
        Cooperation  & \revision{5.0}  & \revision{-1.5} & \cellcolor{lightgray}\revision{0.02}& \revision{-0.82}\\
        \hline        
        Total  & \revision{12.0} & \revision{-1.5} & \cellcolor{lightgray}$<0.01$ & \revision{-0.8}\\
        \hline
    \end{tabular}

    \label{tab:trust}
\end{table}


\subsubsection{NASA TLX Results}
\begin{figure*}[t]
  \centering
  \includegraphics[width=\linewidth]{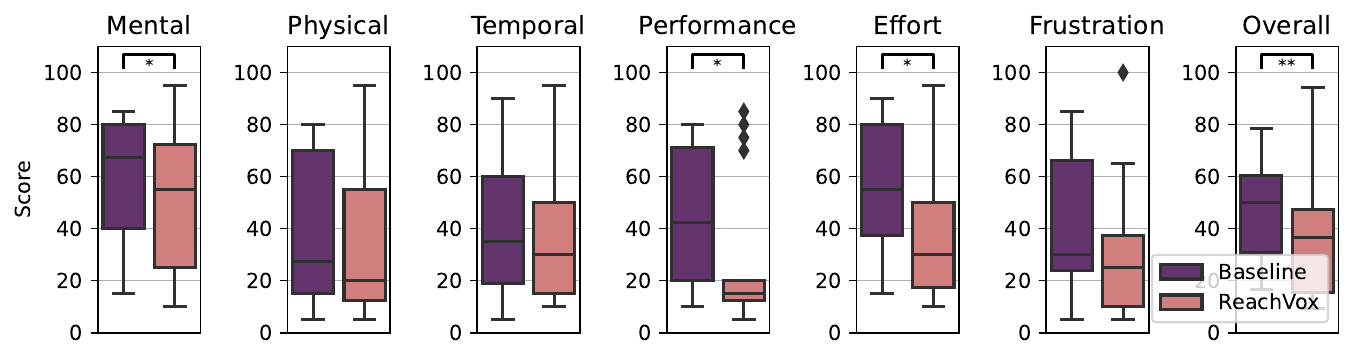}
  \caption{\revision{NASA-TLX results for
   mental demand, physical demand, temporal demand, performance, effort, frustration and overall demand. $*: p < 0.05$. $**: p < 0.01$.}}
  \label{fig:nasatlx}
\end{figure*}

We used the unweighted NASA TLX questionnaire \cite{hart1988development} as an indicator for the perceived workload of participants. The NASA-TLX results indicated a main effect of \revision{\textsc{visualization}} with \revision{mental demand, performance,} effort, and overall demand resulting in significantly better scores for \textsc{ReachVox}, see Table \ref{tab:tlx} and Figure \ref{fig:nasatlx}.

\begin{table}[!htb]
    \centering
    \caption{Wilcoxon signed-rank test results for the NASA TLX with mental demand (Mental), physical demand (Physical), temporal demand (Temporal), performance (Performance), effort (Effort), frustration (Frustration) and overall demand (Overall). md: mean difference. Significant findings are marked in gray.}
    \setlength{\tabcolsep}{6pt}

    \begin{tabular}{|c|c|c|c|c|}
        \hline
        \textbf{Measure} & W & md & p & $r$ \\
           \hline
       Mental  & \revision{111.5} & \revision{17.5} & \cellcolor{lightgray} \revision{0.026} & \revision{0.64} \\
        \hline
        Physical  & \revision{77.0}  & 2.66 & \revision{0.66} & \revision{0.13}\\
        \hline
        Temporal  & \revision{72.0} & $<0.001$ & \revision{0.86} & \revision{0.06}\\
        \hline
        Performance  & \revision{11.5} & \revision{}24.5 & \cellcolor{lightgray} \revision{0.03} & 0.46\\
        \hline
        Effort  & \revision{144.0} & \revision{17.5} & \cellcolor{lightgray}\revision{0.01} & \revision{0.68}\\
        \hline
        Frustration  & \revision{125.5} & \revision{10.0} & \revision{0.08} & \revision{0.47}\\
        \hline
        Overall  & \revision{164.0} & \revision{10.0} & \cellcolor{lightgray}\revision{0.006} & \revision{0.73}\\
        \hline
    \end{tabular}

    \label{tab:tlx}
\end{table}

\subsection{Preferences and Comments}

All \revision{20} participants preferred \textsc{ReachVox}. We clustered the comments of participants from a semi-structured interview following a thematic analysis \cite{guest2011applied}.

\paragraph{ReachVox increased efficiency and confidence.}
Six participants found that using the visualization made the task easier and faster to complete. P5 noted that ``especially during complex tasks (\textsc{hard}) the visualization was very helpful" and P19 commented that ``during the voxel visualization, it was possible to assess whether the configuration made sense purely through visual feedback".\\ 
Five participants found that using visualization increased confidence in their choices, e.g. by increasing the confidence not to damage the workpiece. However, P4 warned that ``the increased confidence and safety could lead to carelessness". Also, two participants found the voxel-visualization to be confusing at first. P5 noted that ``I couldn't immediately understand what I was seeing". 



\paragraph{ReachVox helped to understand robot movement.}
Four participants found that using \textsc{ReachVox} helped them understand the robot movements. P13 noted that ``you could exactly see where the robot would move to". \\
Six participants found that the visualization made it possible to solve the task without moving the robot. Three of those participants found that having the robot in the same setting was a hindrance. P5 noted that ``I didn't need the robot and I had to move it out of the way to see the visualization". P3 found that having no visualization helped to interact with the robot and moved it intuitively. P3 also noted that ``it was easier to just focus on the robot by itself".

\paragraph{The Robot Arm was difficult to operate.}
Seven participants had difficulties when operating the robot arm. P2 noted that they ``found it easier to simulate the robot movement in my head than to try to move the actual robot" and P13 noted that ``it was difficult because the robot would get stuck in a lot of places". Two participants found it exhausting to move the robot and one of those participants experienced motion sickness while operating the robot arm. Three participants also commented that they would have preferred a longer introduction to the robot movement.

\paragraph{The Base Condition led to Trial-and-Error.}
Four participants found that they were more dependent on guessing during the \textsc{base} condition. P15 noted that ``I couldn't understand how the robot would move so I relied mostly on guessing". On starting the base condition, P9 noted that ``I hope I remember the solutions from earlier".

\section{Discussion}

The results of our study indicate that \textsc{ReachVox} outperformed the condition \textsc{base} in terms of objective measures such as task completion time and subjective measures such as usability.

We found that \textsc{ReachVox} supports operators in seeing the effect of moving and rotating the object. For example, rotation is common for improving accessibility on one side of the task space while reducing accessibility on another. By seeing the task space, the robot and the reachability visualization simultaneously supports users to judge the changes of reachability in response to their actions and allowed them to find a spatial configuration that maximizes accessibility faster (as indicated by task completion and number of total changes per trial). In addition, using VR as a wide-angle display enables the user to simultaneously see the object, the task space, and the relationship to the robot in one view and better understand the space.

Interestingly, we did not find a significant difference in robot movement time. Although 6 participants did not move the robot arm at all in \textsc{ReachVox} (one in \textsc{base}), several users moved it out of the way because they found it interfered with visualization.

Robots with multiple joints have complex motions and are hard to imagine intuitively (in many cases, operations focus on the end-effector, but events such as collisions and limitations may occur all along the robot).  We saw that, while the operation of the robot only required positioning the end-effector, several participants struggled with the motion constraints of the robot.  \textsc{ReachVox} concentrated information in a localized visualization with the goal of maximizing visibility (and minimizing occlusions). However, we found that while the participants were trained in using the system, \textsc{ReachVox} was not self-explaining and still confused some users (see comments above).

One could also imagine visualizing further information in the voxel as also proposed by prior work (e.g., \cite{yao2023enhanced}). Expanding the studied welding scenario from point to continuous welding could, for example, benefit from further encoded visualization such as the cumulative joint angle distance between two points in space. In contrast to mere reachability (yes, no), one could also visualize that a point can be reached not only in principle but also that it can be done so without the need to reposition the robot extensively (e.g. starting a weld from left to right, but then requiring repositioning the robot from right to left due to an obstruction).

\revision{Welding may exemplify an application where the operator needs to assess the accessibility of the robot along a whole path or area to enable continuous welding of a part in a continuous manner. This use case is just one of the applications of the robot operation that may extend over a path or an area, from equally covering a part in paint to assembling an object from parts. }

\revision{Such robot applications indicate limitations of many current visualization techniques that often aim at displaying the correctness of a certain robot pose or focus on visualizing occluded space in which a user and robot might collide: many times the designer of the interaction or operator may need to make decisions based on an extended operation space of the robot. A necessary change to correct the relative pose between the robot and its work piece for easy access at the end of a path may affect accessibility at the beginning. It is important to let the user manipulate parameters while viewing the implications on the overall action sequence (in our case, the accessibility of welding points). Current visualizations that incorporate entire actions such as a slew of ghost robots along its path are mostly aiming at visualizing areas that are dangerous for a human co-operator, or to indicate the space occupied by the robot end effector during its motion. These displays tend to obfuscate small details of self-intersection by a part of the robot with the environment or itself. We hope that this work will encourage researchers to look at visualization that includes all the data needed for task success, and that this display is beneficial for efficient work planning.}

Further, ad-hoc situations are not that common on the production floor, as these are designs to avoid surprises but they can be common in remote operations when robots can be in uncontrolled hazards or disaster areas. As more ad-hoc operations are to be used there will be an increased need for further visualizations that better explain the whole task space to the remote decision maker.

One limitation of our work is the use of a naive algorithm to compute the \textsc{ReachVox} visualization, not allowing real-time generation for fine-grained spatial changes of the object of interest but requiring pre-computation. \revision{Hence, as of now, the implementation might not be suitable for situations where real-time updates of the pose between the robot and its workpiece(s) might be required. One example could be the interaction between a physical robot and 3D-reconstructed physical objects being acquired in real-time. However, we are confident that the runtime of the algorithm can be drastically improved through parallelization of the underlying computations. The compute costs could be further minimized by allowing the users to interactively select the regions of interest on a workpiece for which the visualization should be computed.}

Another limitation is the convenience sampling process which led to the inclusion of participants with generally low expertise in robot control. While this allowed us to understand challenges of beginners. Hence, an interesting avenue for future work would be to research the benefits and drawbacks of \textsc{ReachVox} and further visualizations for different target populations such as beginners in robot (tele)operation or seasoned professionals.

\revision{Finally, one can argue that comparing a visualization against no visualization might lead to an expected outcome of the visualization outperforming the baseline. We chose this study design nonetheless as it became evident that existing alternative visualization techniques were unsuitable for the task at hand. Still, we encourage further investigation of visualization alternatives for indicating the reachability in a region of interest. For example, in the future, one potential variation of the \textit{path} visualization could also include a visual reachability encoding along the path segments, which could further minimize visual clutter in the scene.}






\section{Conclusion and Future Work}
In this paper, we presented \textsc{ReachVox}, a visualization approach to aid robot motion planning tasks in VR. Based on the idea of reachability maps, we visualize the reachability of the volume around an object of interest with reduced visual clutter compared to alternative techniques. Our user study indicated that \textsc{ReachVox} outperformed the \textsc{base} condition in most objective and subjective measures and that participants clearly favored \textsc{ReachVox}. In future work, we want to include more efficient motion planning algorithms and parallelization to study the real-time generation capabilities for local reachability volumes, \revision{and study use cases beyond stationary robotic arms such as humanoid robots or manipulators on motion platforms. Specifically, we plan to study \textsc{ReachVox} in a joint human-robot nursing scenario in which a manipulator on a motion platform supports a human in a joint gripping procedure (e.g. the robot holding a leg of a patient while the nurse changes a bandage). In this context, one could study the extension of \textsc{ReachVox} to visualize the \textit{joint} reachability of a region of interest as well as further parameters such as applied gripping force at a contact point}.


\bibliographystyle{abbrv-doi}

\balance
\bibliography{template}
\end{document}